\documentclass[conference]{IEEEtran}

\ifCLASSINFOpdf
\usepackage[pdftex]{graphicx}
\else
\fi

\usepackage{balance}
\usepackage{cite}
\usepackage{graphicx}
\usepackage{float}
\usepackage{amsmath}
\usepackage[utf8]{inputenc}
\usepackage[final]{pdfpages}
\usepackage{pdfpages}
\usepackage{lipsum}
\usepackage{textcase}
\usepackage{url}
\usepackage{amsmath,esint}
\usepackage{epstopdf}
\usepackage{array}
\usepackage{color}
\usepackage{subcaption}

\usepackage{bm}

\DeclareRobustCommand{\uvec}[1]{{%
  \ifcsname uvec#1\endcsname
     \csname uvec#1\endcsname
   \else
    \bm{\hat{\mathbf{#1}}}%
   \fi
}}

	\newcommand\blfootnote[1]{%
		\begingroup
		\renewcommand\thefootnote{}\footnote{#1}%
		\addtocounter{footnote}{-1}%
		\endgroup
	}
\newcolumntype{P}[1]{>{\centering\arraybackslash}p{#1}}
\newcolumntype{M}[1]{>{\centering\arraybackslash}m{#1}}

\pdfoutput=1

\begin{document}

\title{Temporal and Spatial Characteristics of mmWave Propagation Channels for UAVs}

\author{
\IEEEauthorblockN{Wahab Khawaja, Ozgur Ozdemir, and 
Ismail Guvenc}
\IEEEauthorblockA{Department of Electrical and Computer Engineering, North Carolina State University, Raleigh, NC}
Email: \{wkhawaj, oozdemi, iguvenc\}@ncsu.edu
	
}

\maketitle
\blfootnote{This work has been supported in part by NSF under the grant CNS-$1453678$ and by NASA under the Federal Award ID number
NNX17AJ94A.}

\begin{abstract}
Unmanned aerial vehicles~(UAVs) are envisioned to be an integral part of future 5G communication systems. The agile nature of UAVs for serving users at different locations can help to dynamically optimize coverage and quality-of-service (QoS) in future networks. 
In this work, we explore the small scale temporal and spatial characteristics of mmWave air-to-ground~(AG) line-of-sight~(LOS) propagation channels at $28$~GHz in different environmental scenarios: dense-urban, suburban, rural, and over sea using omni-directional antennas employing Wireless InSite ray tracing software. We classify the received multipath components~(MPCs) into persistent and non-persistent components. The small scale temporal and spatial characteristics of the AG propagation channel are found to be dependent on the scatterer properties: number, distribution, and geometry. Additionally, clustering of MPCs in the time and spatial domain for different environments is found to be dependent on the scatterer properties and receiver sensitivity. When the height of the UAV is comparable to the height of the  scatterers, we observe large temporal and angular spreads.


\end{abstract}

\IEEEpeerreviewmaketitle

\section{Introduction}
The use of civilian unmanned aerial vehicles~(UAVs) has seen an exponential growth in recent years~\cite{Tractica}. 
Ensuring reliable broadband connectivity at UAVs carries critical importance for wide range of usecases. 
Use of UAVs to be served by existing 4G LTE networks has already been getting extensive  attention recently from academia and standardization organizations~\cite{LTEUAV, qualcom&report:2017,3GPPUAV}. 5G technology utilizing mmWave frequencies will allow even higher throughput services with UAVs~\cite{uav_previous,xiao2016enabling,rupasinghe2017non}.
To our best knowledge, mmWave air-to-ground~(AG) communication channel in different deployment environments is not studied extensively in the literature to date.  

The available literature for higher frequency AG propagation channel covers mostly the long distance satellite links~\cite{Satellite1,Satellite2,Satellite3}. 
For relatively short distance mmWave communications using UAVs, the propagation channel characteristics will differ significantly. A major difference is due to the contribution of scatterers on the ground. In long-range AG propagation links, the effect of scatterers will be smaller when compared to short distance AG links, especially in a dense-urban area. Additionally, in case of short distance AG propagation channel, the atmospheric effects are relatively lower. 
In~\cite{Literature1}, an overview is provided for current channel models being developed for terrestrial $5$G networks using mmWave frequencies. A similar study on the future channel models using mmWaves is presented in \cite{Literature2}, while~\cite{khawaja2018survey} provides a comprehensive survey on AG propagation channels for UAVs. 

This paper is an extension of \cite{uav_previous}, where, time dispersion and large scale propagation characteristics at $28$~GHz are studied for different environmental scenarios. Here, small scale characteristics of the mmWave propagation channel in both time and spatial domains are analyzed (in contrast to~\cite{uav_previous} which considers only time domain characteristics) in dense-urban, suburban, rural and over sea scenarios using Wireless InSite ray tracing software. 
The analysis of the mmWave AG propagation channel reveals that the received multipath components~(MPCs) can be grouped into persistent (comprising of LOS and ground reflected component~(GRC)) and non-persistent (comprising of all other non-LOS (NLOS) components). The characteristics of persistent components are mainly dependent on the geometry of the setup and are predictable for a given trajectory of the UAV. On the other hand, statistics of the non-persistent components follow a random process along the UAV trajectory and are dependent on the properties of the terrain cover in addition to the geometry of the setup. 


\section{Ray Tracing Simulations}\label{Section:Ch_simulations}
The following section describes the mmWave channel simulation setup for different propagation environments using the ray tracing simulation tool. 

\subsection{Terrain Parameters for Ray Tracing Simulations}
We utilize the Remcom Wireless InSite software to estimate the electromagnetic radio wave propagation through different environments. In case of over sea scenario, a layer of sea water of $10$~m height is considered on the top of the ground terrain of size $10~{\rm km}\times10~{\rm km}$. The sea surface is calm without any turbulence and the effect of the sea water is only in the form of additional attenuation due to refraction of the electromagnetic waves traveling from one medium to the other. This mainly results in the reduction of received power of the GRC. 
For ground based AG propagation channel, dense-urban, suburban, and rural scenarios are created on the same size ground terrain as used in over sea scenario. In all the ground based cases, wet earth is used as ground terrain material. 

The four different environmental scenarios are classified based on the number, distribution and dimensions of the buildings. In case of dense-urban scenario, the buildings are closely packed with complex structures and higher heights, whereas in case of suburban and rural, the buildings are scattered sparsely in the environment with lower heights and simpler structures, as specified in Table~\ref{Table:Table_RT}. The material used for all the buildings is frequency sensitive concrete at $28$~GHz. The dimensions and placement of the buildings are arbitrary. This ensures to avoid over-fitting to a specific propagation area. The effects of other terrain covers such as lamp posts and sign boards are considered to be negligible due to smaller dimensions as compared to buildings. Foliage area consisting of branches and leaves of trees is also considered in the suburban and rural scenarios but has negligible effect on the overall simulations.

\begin{table}[!t]
	\begin{center}		
		\caption{Deployment parameters for four different ray tracing simulation environments.}\label{Table:Table_RT}
        	\begin{tabular}{|P{2cm}|P{2.5cm}|P{2.8cm}|}
			\hline
			\textbf{Scenario}&\textbf{Building height (m)}&\textbf{Number of buildings}\\
			\hline
			Over sea &-& - \\
            \hline
            Rural & 4-8&10\\
            \hline
            Suburban & 4-30&20 \\
            \hline
            Urban & 70-180&100 \\
            \hline
            
		  \end{tabular}
		\end{center}
        \vspace{-4mm}
 \end{table}

\subsection{Propagation Scenarios for Ray Tracing}
A sinusoidal sounding signal is used at a center frequency of $28$~GHz. Vertically polarized half-wave dipole antennas are used both at the transmitter and at the receiver in the simulations that have an omni-directional pattern in the azimuth direction. The half power beam width of antenna radiation pattern in the E-plane is $90^\circ$, whereas the first null beam width in the E-plane is $180^\circ$. The signal is transmitted with $30$~dBm power and the receiver sensitivity is limited to $-110$~dBm. Receiver sensitivity threshold selection is critical for mmWave communications due to higher path losses. A receiver with higher sensitivity will be able to capture higher number of the MPCs that increases the granularity of the channel analysis.  

\begin{figure}[t]
	\centering
	\includegraphics[width=\columnwidth]{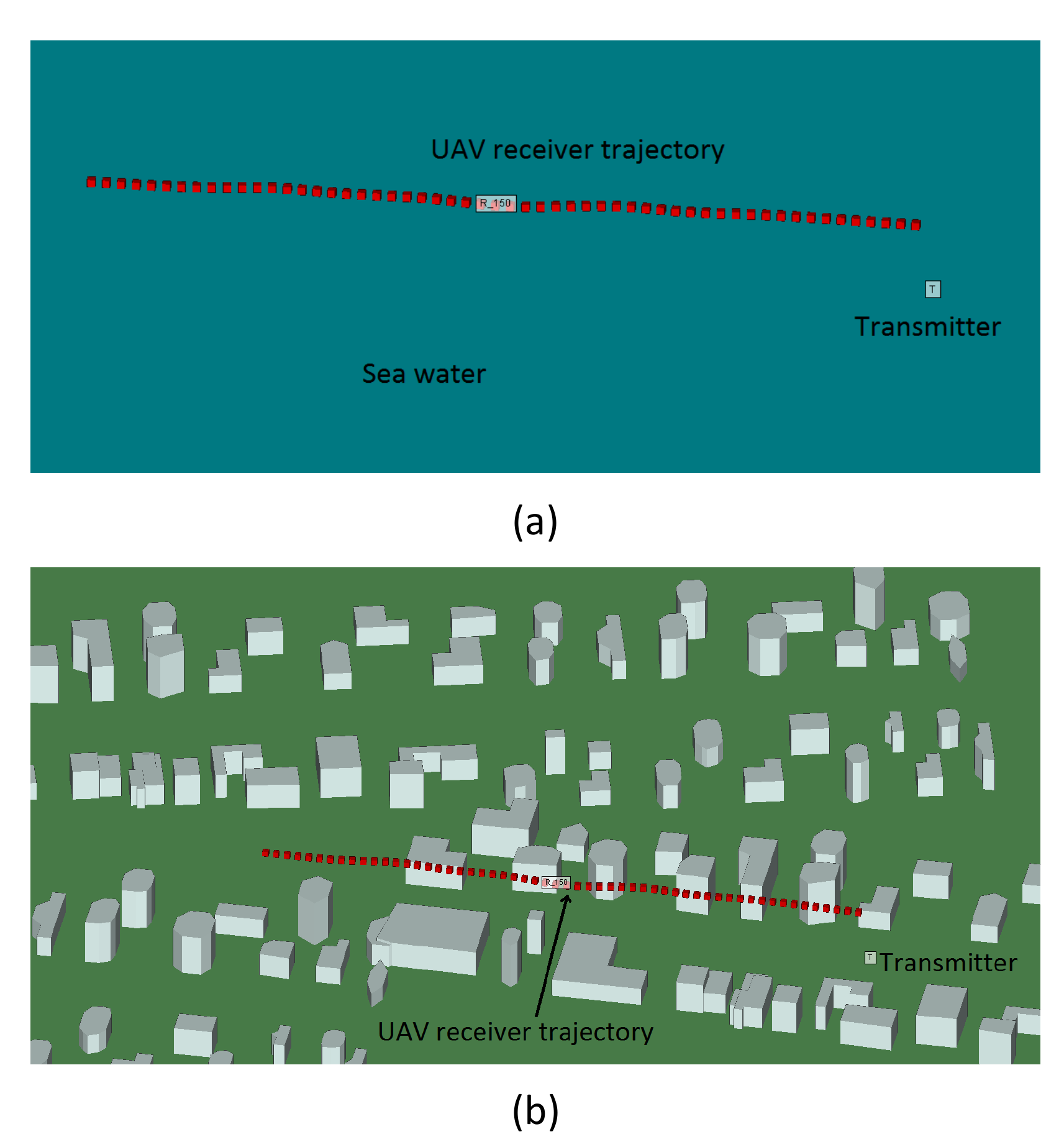}
	    \caption{Simulation scenario of AG propagation in Remcom Wireless InSite. The transmitter is placed at a height of 2~m from sea surface and ground, whereas the receiver is placed on the UAV at a height of $150$~m that follows the trajectory illustrated. (a) Over sea scenario, (b) Dense-urban scenario. 
}\label{Fig:setup}\vspace{-4mm}
\end{figure}

The transmitter is placed on the ground at a height of $2$~m above the ground, whereas the receiver is placed on the UAV at four different heights from ground: $2$~m, $50$~m, $100$~m and $150$~m. The horizontal distance of transmitter from the receiver trajectory starting point is $40$~m. Each point on the UAV trajectory is a point receiver with uniform distance among them. The velocity of the UAV is fixed at $15$~m/s for all the scenarios. Due to small velocity considered, the changes in the roll angles will be negligible, while the UAV to ground transmitter distance is monotonically increasing as UAV moves on an approximately straight line. The UAV trajectory is aligned to the transmitter in the azimuth plane without any sharp turns as shown in Fig.~\ref{Fig:setup}. This ensures a direct LOS path at all the times along the link distance. The length of the UAV trajectory is limited to $1.2$~km due to higher attenuation losses at larger link distances.

\subsection{Channel Impulse Response} \label{Section:CIR}
The CIR considering both temporal and spatial characteristics can be represented as follows:
\begin{align}
h(t,\boldsymbol{\tau},{\boldsymbol\theta}, {\boldsymbol\phi}) =& \sum_{l = 1}^{L(t)}\alpha_{l}(t)\exp\big(j\psi_{l}(t)\big)\delta\big(t-\tau_{l}(t)\big)\nonumber\\ &\times\delta\big({\theta} - {\theta}_{l}(t)\big)\delta\big({\phi} - {\phi}_{l}(t)\big), \label{Eq:CIR}
\end{align}
where $L(t)$ is the number of multipath components at time instant $t$, $\alpha_{l}(t)$, $\tau_l(t)$, $\psi_{l}(t)$ are the amplitude, time of arrival~(TOA), and phase of the $l^{\rm th}$ MPC at time instant $t$, whereas
${\theta}_{l}(t)$ and ${\phi}_{l}(t)$ are the direction of departure~(DOD) and direction of arrival~(DOA) at the transmitter side and receiver side, respectively, of the $l^{\rm th}$ MPC at time instant $t$. The TOA, DOD, and DOA parameters can further be grouped under the vectors $\boldsymbol{\tau}=[\tau_1,...,\tau_{L(t)}]$, $\boldsymbol{\theta}=[\theta_1,...,\theta_{L(t)}]$, $\boldsymbol{\phi}=[\phi_1,...,\phi_{L(t)}]$, respectively. Both the DOD and DOA for the $l^{\rm th}$ MPC are composed of elevation and azimuth elements, which can be represented as $\theta_{l}(t)=\theta_l^{\rm (e)}(t) + j\theta_l^{\rm (a)}(t)$ and ${\phi}_{l}(t)=\phi_l^{\rm (e)}(t)+ j\phi_l^{\rm (a)}(t)$.

\begin{figure}[!t]
	\centering
	\includegraphics[width=0.75\columnwidth]{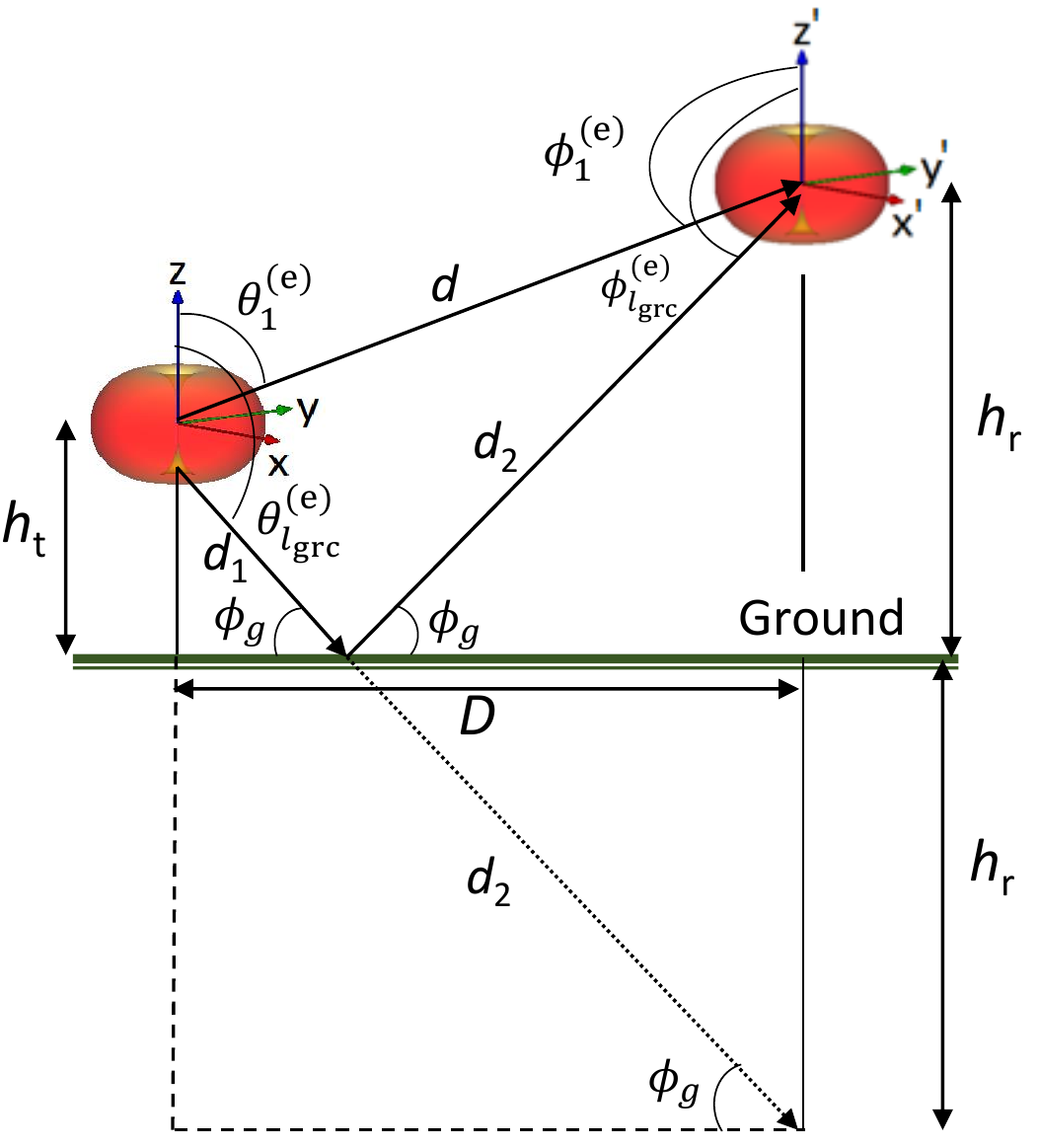}
	\caption{Persistent components for the AG propagation channel in the elevation plane with LOS and GRC links. Antenna radiation patterns for the ground transmitter and the UAV receiver are also illustrated. The UAV and ground transmitter heights are denoted by $h_{\rm r}$ and $h_{\rm t}$ with $h_{\rm r}\geq h_{\rm t}$. The MPC index for the GRC path is denoted by $l_{\rm grc}$.}\label{Fig:Persistence_comp}\vspace{-4mm}
\end{figure}

We will further group the paths into persistent and non-persistent components. The persistent components consist of the LOS and the GRC paths as illustrated in Fig.~\ref{Fig:Persistence_comp} which also shows the elevation DOD and DOA of both components. Due to a clear LOS path considered for the specific scenario in our simulations as in Fig.~\ref{Fig:setup}, the LOS and GRC components will always be present. The non-persistent components follow a given birth and death process. This constitutes a random process along UAV route dependent on the terrain cover characteristics represented as a random vector.  
Letting $l=1$ corresponds to the always-present LOS component, and the power of the LOS component $\alpha_{1}$ along the UAV trajectory can be represented as:
\begin{equation}\label{Eq:Amplitude}
P_{1} (t) = \alpha^2_{\rm ref}\beta^{\gamma}g(\phi_{\rm e}), 
\end{equation}
where $\alpha_{\rm ref}$ is the amplitude of the MPC at a reference position, $\beta = \big(\frac{4\pi d}{\lambda}\big)^{-1}$, $\gamma = 2$ for free space, $d$ is the slant range, and $g(\phi_{\rm e})\in [0,1]$ is a monotonically decreasing function of $\phi_{\rm e}$, in order to characterize the antenna radiation pattern misalignment (due to \emph{donut-shaped} radiation pattern) between the transmitter and the receiver as illustrated in Fig.~\ref{Fig:Persistence_comp}.
The phase of the LOS component is given as $\frac{2\pi fd}{c}$, where $\frac{d}{c}$ is the TOA of the LOS component. The DOA of the LOS component in the elevation plane depends on the height of the UAV, and it decreases along the UAV trajectory for a given fixed UAV height as shown in Fig.~\ref{Fig:Persistence_comp}. 

For the GRC, we have similar expression as in (\ref{Eq:Amplitude}), but it is additionally multiplied with the modulus of ground reflection coefficient~\cite{GRC}. This coefficient will be dependent on the ground characteristics and grazing angle represented as $\phi_{\rm g} = \arctan\big(\frac{h_{\rm t}+h_{\rm r}}{D}\big)$ 
in Fig.~\ref{Fig:Persistence_comp}. The phase of the GRC is composed of two components, one is dependent on the distance between the transmitter and the other is due to reflection from the ground surface. The TOA of the GRC is given as $\frac{d_1+d_2}{c}$ from Fig.~\ref{Fig:Persistence_comp}. The elevation angle for the GRC is higher than the LOS component, and it decreases along the UAV trajectory as the horizontal distance $D$ increases for a given UAV height. 

For non-persistent components, the amplitude, phase, TOA and DOA of the MPCs can be characterized by a random process whose statistical characteristics will be dependent on the scatterer number, dimensions, distribution, and material for a given simulation setup. In Sections~\ref{Sec:Power}-\ref{Sec:Spatial}, building on the framework in this section, we will present ray tracing  results for the UAV trajectories as in Fig.~\ref{Fig:setup}, and will discuss power variation, temporal characteristics, and spatial characteristics of UAV AG propagation channels.  

\section{Power of Multipath Components}\label{Sec:Power}

\begin{figure}[!t]
	\centering
	\includegraphics[width=\columnwidth]{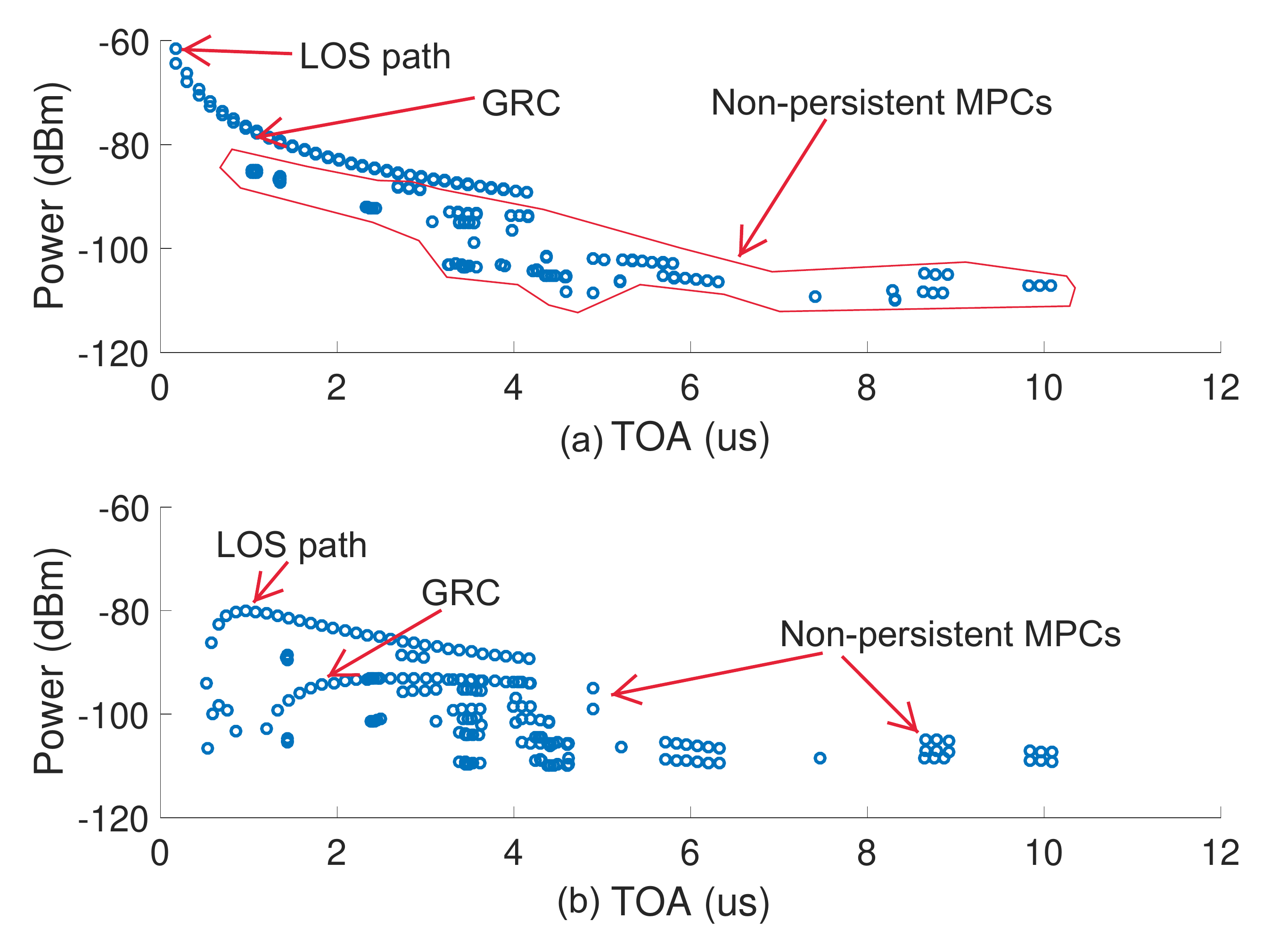}
	\caption{Received power of MPCs plotted against the TOA of MPCs over the UAV trajectory for dense urban scenario at a UAV height of (a), and $2$~m (b) $150$~m scenario.}\label{Fig:Received_power_TOA_urban}\vspace{-3mm}
\end{figure}

First, consider that both the GS and the UAV are at the same height as in Fig.~\ref{Fig:Received_power_TOA_urban}(a). Then, there is no antenna misalignment in the vertical direction and the factor $g(\phi_{\rm e})$ in (\ref{Eq:Amplitude}) will be one, and the only loss observed is the free space path loss for the LOS component. Therefore, the power of a MPC decreases monotonically as the UAV travels on its trajectory for all the MPCs. For the GRC, the amplitudes will be slightly smaller than the LOS component due to ground reflection coefficient as discussed in Section~\ref{Section:CIR}. 

In Fig.~\ref{Fig:Received_power_TOA_urban}(b) at a UAV height of $150$~m, we observe that the received power for a MPC tends to initially  increase as the UAV moves away from the transmitter, and then decrease for both LOS and GRC components. This is due to higher misalignment losses in the start (for small $D$) given by $g(\phi_{\rm e})$, that decreases along the UAV trajectory. This results in an initial increase in the  received power, but eventually the free space path loss dominates. For GRC, the received power reduction due to misalignment is higher due to ground reflection coefficient~\cite{GRC}.   

\begin{figure}[!t]
	\centering
    \includegraphics[width=\columnwidth]{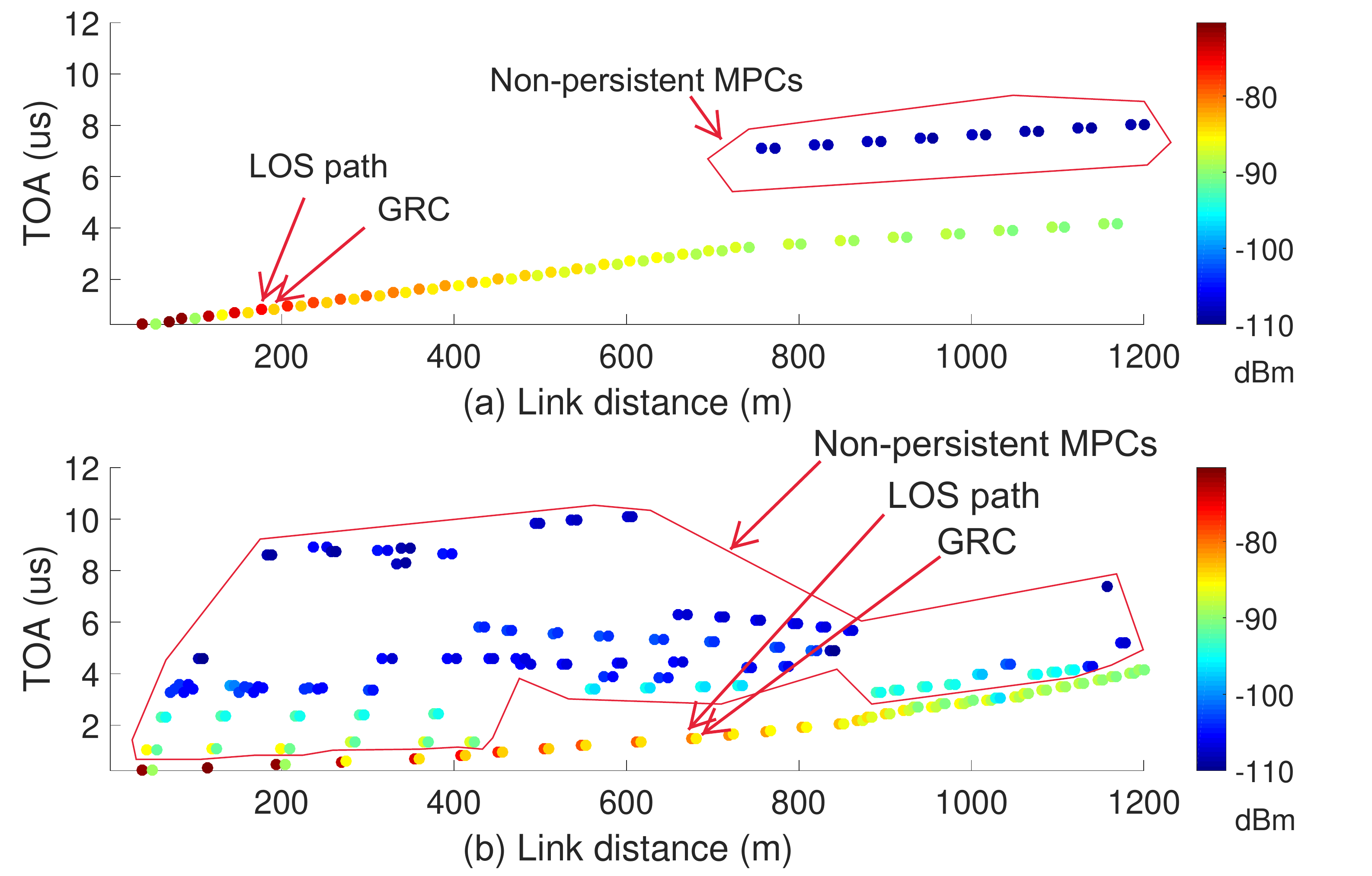}
	\caption{TOA of MPCs over the UAV trajectory at a UAV height of $50$~m for (a) Suburban scenario (b) Dense-urban scenario.} \label{Fig:TOA_RX_50m_suburban_urban}
    \vspace{-3mm}
\end{figure}

\section{Temporal Analysis of mmWave UAV  Channels}\label{Sec:Temporal}

\begin{figure}[!t]
	\centering
	\includegraphics[width=0.97\columnwidth]{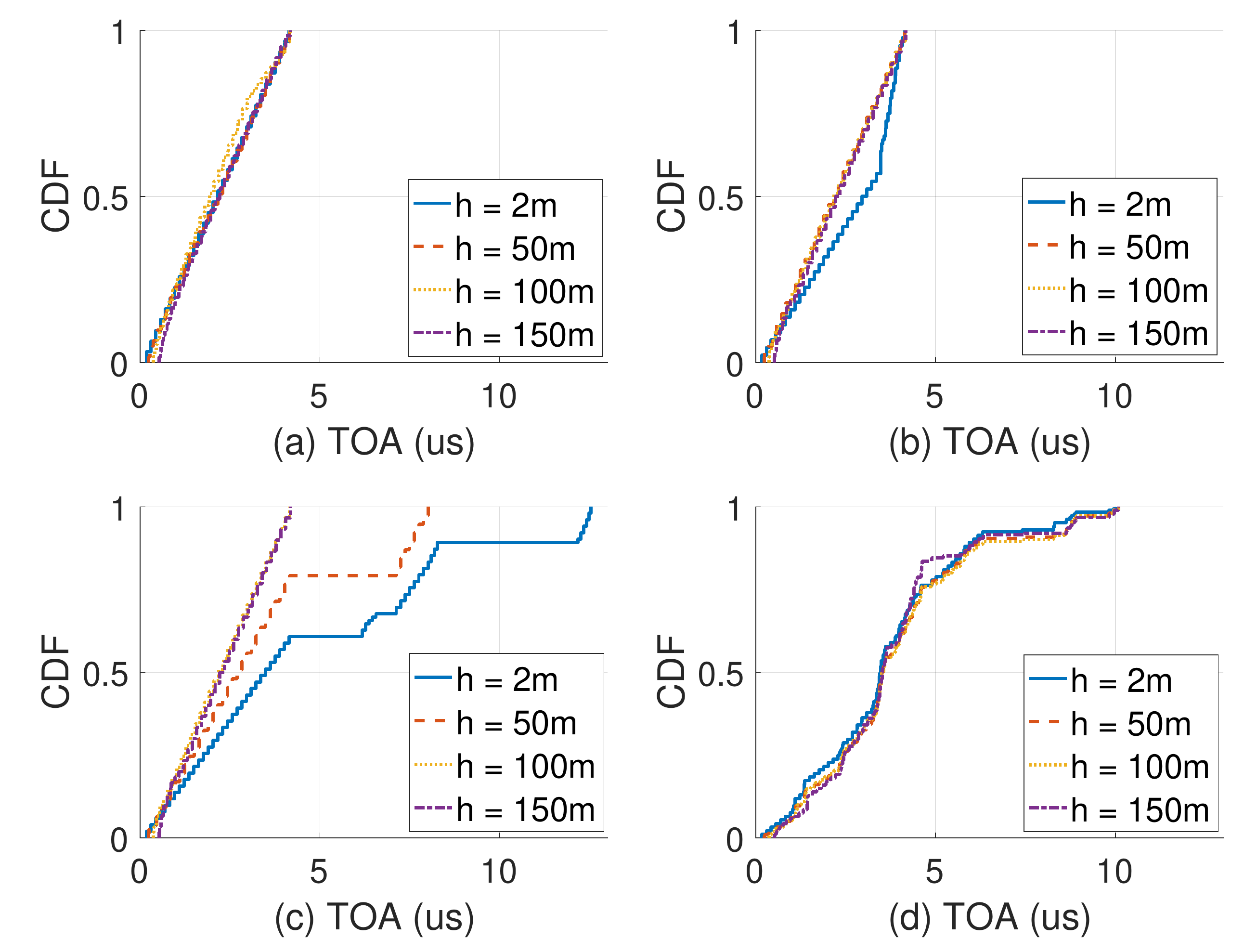}
	\caption{CDF of TOA of MPCs at $28$~GHz for different propagation environments, (a) Over sea, (b) Rural, (c) Suburban, (d) Dense-urban.}\label{Fig:CDF_TOA_28GHz}\vspace{-3mm}
\end{figure}

In this section, we will study the temporal characteristics of the UAV mmWave propagation channel for the scenario in Fig.~\ref{Fig:setup}. First, in Fig.~\ref{Fig:TOA_RX_50m_suburban_urban}, the TOA and the signal strength (through the color bar) of the MPCs arriving at the UAV are studied as a function of link distance for suburban and urban environments. 
Since the UAV to transmitter distance increases monotonically for the particular example in Fig.~\ref{Fig:setup}, the TOA of the persistent MPCs are expected to change approximately linearly as well along the UAV trajectory. Then, Fig.~\ref{Fig:CDF_TOA_28GHz} studies the cumulative distribution function (CDF) of the TOAs of MPCs in four different environments. 

In Fig.~\ref{Fig:TOA_RX_50m_suburban_urban}(a), a suburban propagation scenario at a UAV height of $150$~m is presented, where we can observe that there is only the LOS and the GRC links present. The TOA of the LOS component is varying approximately linearly as well as the received power, whereas the GRC also shows a similar trend but at a different rate. Initially, the received power is smaller and then increases, and later follows a decreasing trend. The decrease in the start can be described by the higher elevation angle difference resulting in additional loss due to misalignment between the transmitter and receiver antennas. However, with increasing link distance, the misalignment in the elevation plane decreases, resulting in increase in the received power. The received power  decreases later (as was discussed in Fig.~\ref{Fig:Received_power_TOA_urban}) due to free space loss in case of LOS component and ground reflection coefficient in addition to free space path loss for the GRC.  

Fig.~\ref{Fig:TOA_RX_50m_suburban_urban}(b) shows, on the other hand, results in a dense-urban scenario at a UAV height of $150$~m. A similar behavior to Fig.~\ref{Fig:TOA_RX_50m_suburban_urban}(a) is observed for the persistent components in dense urban scenario, except that there are additional non-persistent MPCs due to scattering from densely deployed buildings. This results in overall higher TOA as compared to suburban scenario. The TOA of these MPCs are randomly scattered, and are dependent on the terrain cover properties vector. These non-persistent MPCs follow a given birth death process along the trajectory of the UAV~\cite{Survey_Matolak}. 

In Fig.~\ref{Fig:CDF_TOA_28GHz}, CDFs of the TOAs for the MPCs are shown for different UAV heights and propagation environments. We can generally observe an approximately linear response in the CDF plots indicating the influence of persistent components. 
In Fig.~\ref{Fig:CDF_TOA_28GHz}(a), the TOA of MPCs for over sea scenario is plotted for four UAV heights. We approximately have linear CDFs except at $150$~m,  owing to refraction loss from the sea water in addition to the height of the UAV. 
For the rural scenario in Fig.~\ref{Fig:CDF_TOA_28GHz}(b), the response is similar as in case of over sea scenario. This is due to the fact that there are very few buildings with lower heights that marginally reflect the transmitted waves. The scattering effect is only visible for a UAV height of $2$~m. As the height of the UAV increases, the reflected MPCs from the few small buildings will not reach the UAV resulting in mostly persistent components.

For the suburban scenario in Fig.~\ref{Fig:CDF_TOA_28GHz}(c), the CDF for UAV heights of $100$~m and $150$~m show only the effect of persistent components as observed in the over sea and rural scenario. Whereas, at UAV heights of $2$~m and $50$~m, we observe higher variance in the TOA of the MPCs due to their interaction with the buildings of comparable or higher heights with respect to the UAV height. 
For the dense-urban scenario in Fig.~\ref{Fig:CDF_TOA_28GHz}(d), we observe even higher variance in the TOA of the MPCs as compared to all other propagation environments. This is due to comparable or higher heights of large number of buildings with respect to the UAV heights. This results in significant number of non-persistent MPCs reaching the UAV in addition to the persistent components as observed in Fig.~\ref{Fig:TOA_RX_50m_suburban_urban}(b). In this case, the TOAs of the MPCs are affected approximately equally at all the UAV heights.  


\section{Spatial Analysis of mmWave UAV Channels}\label{Sec:Spatial}

In this section, spatial characteristics (in particular, DOAs and DODs) of the MPCs at the transmitter and receiver are analyzed for different propagation environments.

\subsection{Direction of Arrival of MPCs}
The DOA of any MPC can be determined at the receiver by the knowledge of angles that it forms in the elevation and azimuth planes.

\subsubsection{Azimuth Direction of Arrival}
The use of omni-directional antennas provides the ability to capture the MPCs from all directions in the azimuth plane~($0^\circ$-$360^\circ$). Since the GS is fixed on the ground~(no sideways movement with respect to the UAV trajectory that is approximately straight), there will be negligible change in the angles in the azimuth plane for the persistent components considering the trajectory in Fig.~\ref{Fig:setup}. On the other hand, for non-persistent components the DOAs will be mainly dependent on the random vector of terrain cover. 

\begin{figure}[!t]
	\centering
    \includegraphics[width=\columnwidth]{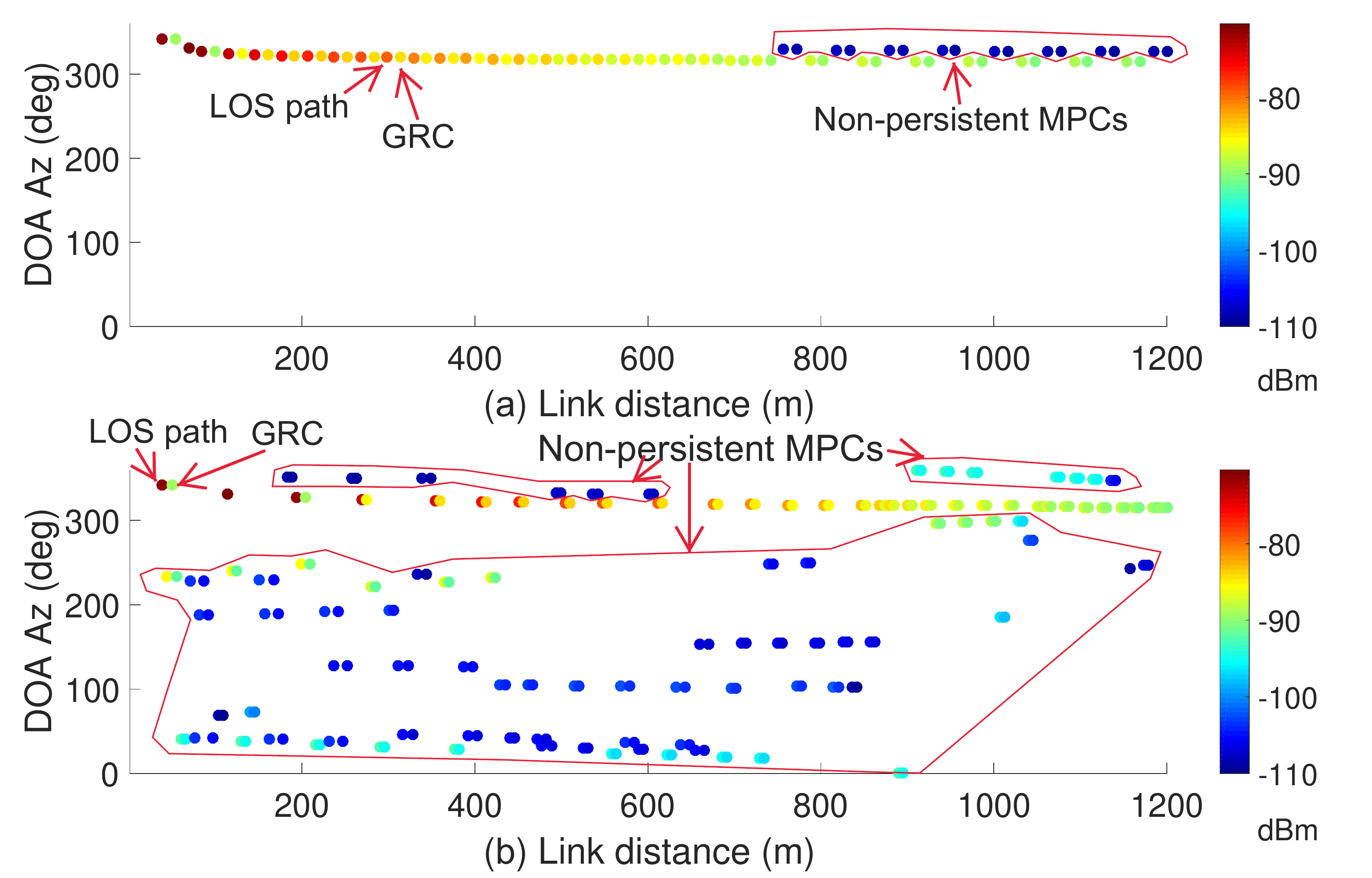}
	\caption{DOA in the azimuth plane for the UAV trajectory at a height of $50$~m for (a) Suburban (b) Dense-urban scenario.} \label{Fig:DOA_Az_RX_50m_suburban_urban}\vspace{-3mm}
\end{figure}


Fig.~\ref{Fig:DOA_Az_RX_50m_suburban_urban} shows the DOAs of the MPCs at UAV height of $50$~m for suburban and dense-urban scenarios. It can be observed that the DOA remains the same for both the suburban and dense-urban scenarios for the persistent components observed as a continuous line with decreasing received power along the link distance.  The value of the DOA is dependent on the reference selected, as well as the position of the transmitter and the receiver in the azimuth plane. 
There are non-persistent MPCs at respective azimuth DOAs that can be observed in addition to the persistent component in Fig.~\ref{Fig:DOA_Az_RX_50m_suburban_urban}(a) and more randomly scattered ones in the dense-urban scenario in Fig.~\ref{Fig:DOA_Az_RX_50m_suburban_urban}(b). Note that even though the received powers are typically very weak for the non-persistent MPCs, with high enough transmit and receive beamforming gains, it may still be possible to compensate for the high path loss and communicate over the NLOS MPCs (e.g. due to blockage effects). 

\begin{figure}[!t]
	\centering
	\includegraphics[width=0.97\columnwidth]{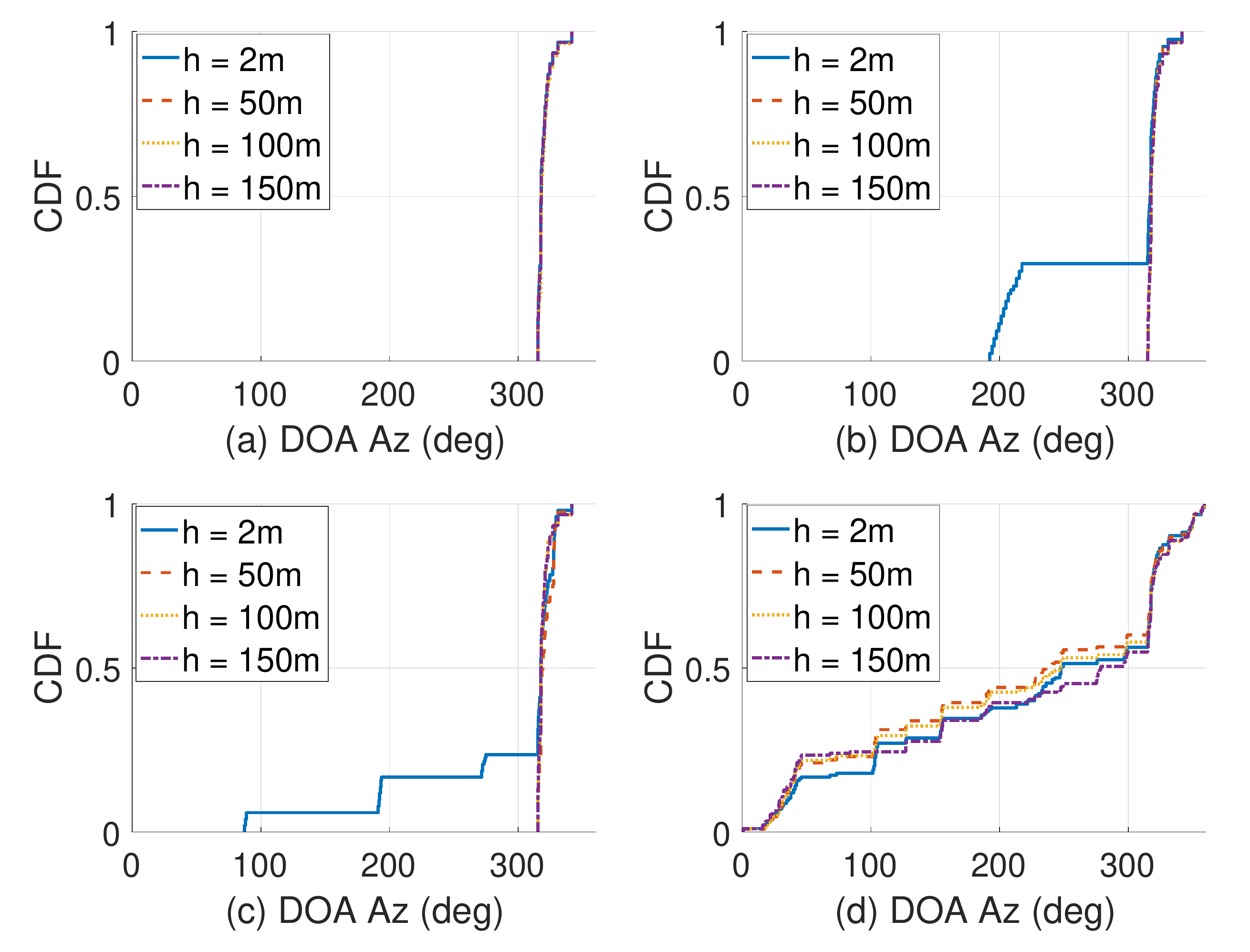}
	\caption{CDF of DOA of MPCs at $28$~GHz in the azimuth plane for different propagation environments, (a) Over sea, (b) Rural, (c) Suburban, (d) Dense-urban.}\label{Fig:CDF_AOA_Az}\vspace{-5mm}
\end{figure}

The CDF of the DOAs of the MPCs in the azimuth plane for different propagation environments at respective UAV heights are shown in Fig.~\ref{Fig:CDF_AOA_Az}. The DOA of the MPCs for over sea scenario in Fig.~\ref{Fig:CDF_AOA_Az}(a) depicts approximately constant probability of occurrence of MPCs at a reference angle of around $325^\circ$ (the value is dependent on the orientation of the GS and the UAV trajectory). This angle corresponds to the persistent components that have the similar angle in the azimuth plane from the GS. 
In case of rural scenario shown in Fig.~\ref{Fig:CDF_AOA_Az}(b), we have similar behavior as observed in the over sea scenario with the exception that at $2$m, we observe additional variance in the azimuth DOA with the range extending to $190^\circ$. This is due to the MPCs from the scatterers of comparable or higher heights as compared to UAV height of $2$~m. Similar observation can be made for the suburban scenario in Fig.~\ref{Fig:CDF_AOA_Az}(c). 

For the dense-urban scenario, on the other hand, we have higher variance of the azimuth DOA at all heights of the UAV between $[0^\circ-360^\circ]$. This high variance is imparted by the non-persistent MPCs that have randomly distributed DOAs in the azimuth plane due to reflections from multiple scatterers around the UAV trajectory, resulting in higher angular spreads. This also implies that if the LOS/GRC paths are blocked for a UAV, there can be several alternative NLOS paths to communicate in urban environments due to rich scattering.

\begin{figure}[!t]
	\captionsetup[subfigure]{labelformat=empty}
	\begin{subfigure}{0.5\textwidth}	
	\centering
    \includegraphics[width=\columnwidth]{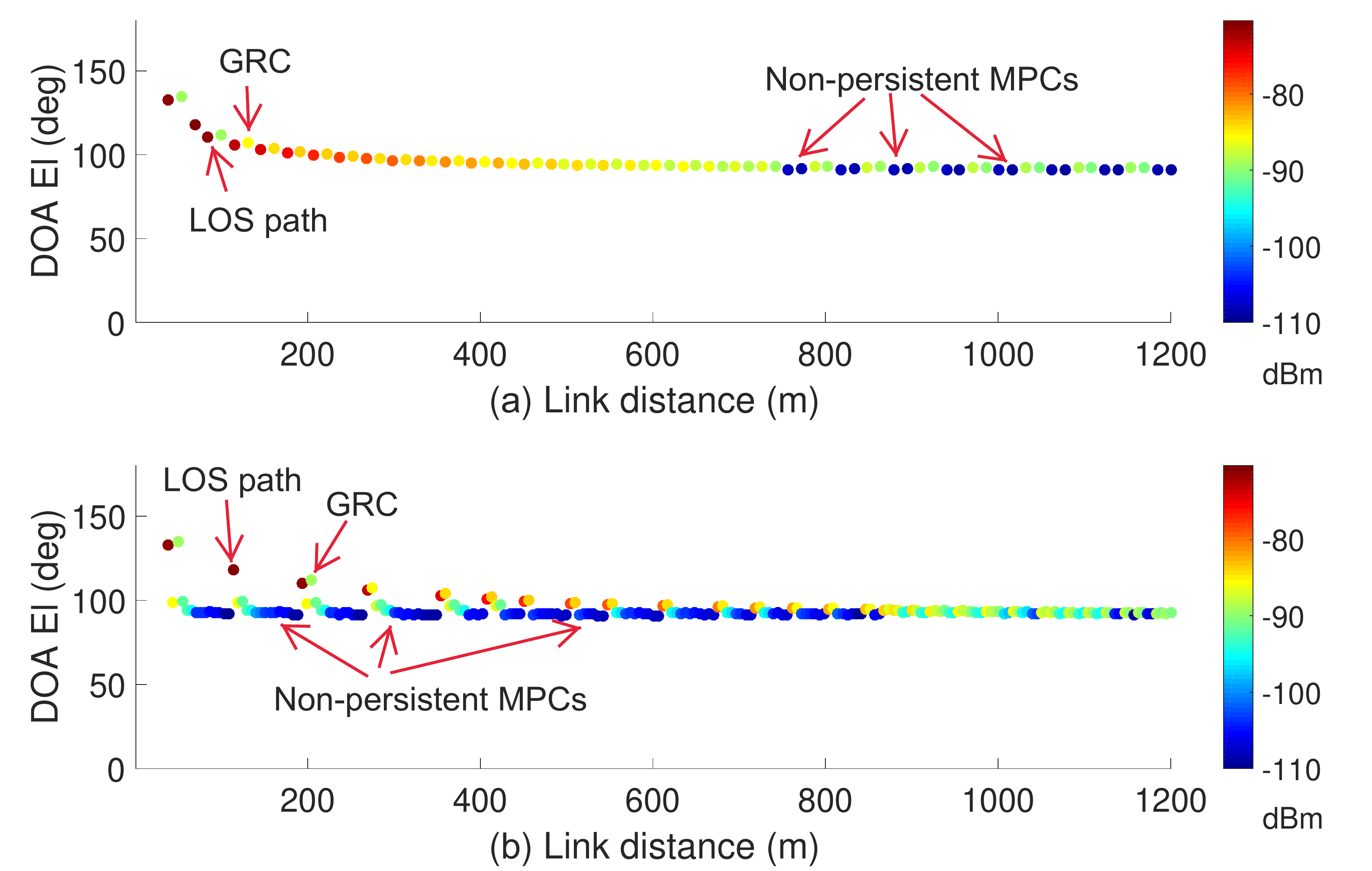}
	\caption{}\vspace{-4mm}
    \end{subfigure}
	\begin{subfigure}{0.5\textwidth}
	\centering
    \includegraphics[width=\columnwidth]{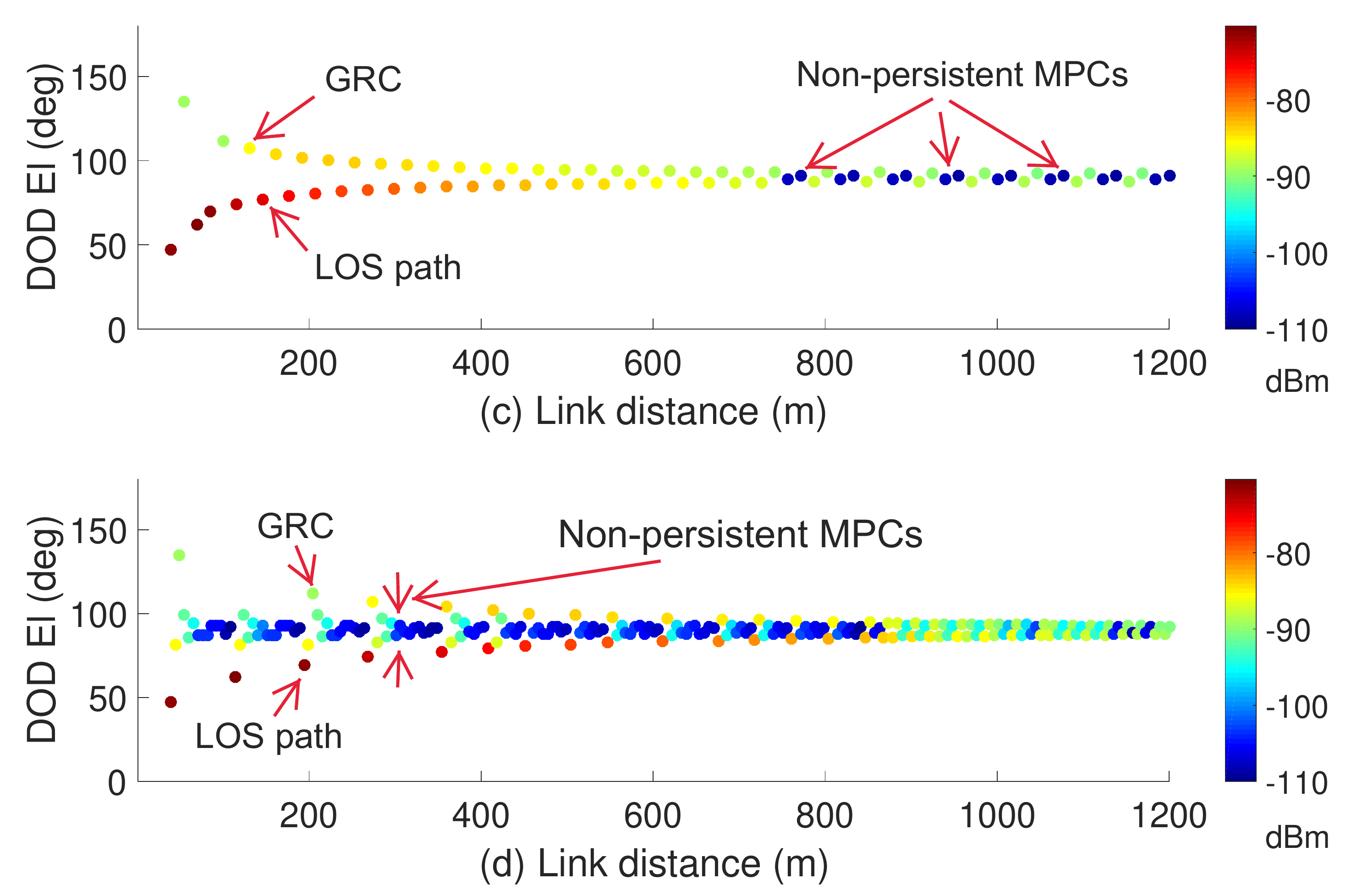}
    \end{subfigure}
    \caption{DOA in the elevation plane for the UAV trajectory at a height of $50$~m for (a) Suburban, (b) Dense-urban scenario; DOD in the elevation plane for the UAV trajectory at a height of $50$~m for (c) Suburban, (d) Dense-urban scenario.} \label{Fig:AOA_AOD_El_overall_50m_suburban_urban}            
       \vspace{-4mm}
\end{figure}

%


\subsubsection{Elevation Direction of Arrival} \label{Subsection:Elevation_AoA}

The elevation angle of persistent and non-persistent components are dependent on the height of the UAV. Additionally, for a given UAV height, the trajectory height remains constant over the link distance. A plot of elevation DOA of MPCs for suburban and dense-urban scenarios are shown in Fig.~\ref{Fig:AOA_AOD_El_overall_50m_suburban_urban}(a)-(b). It can be observed that the elevation angle varies along the UAV trajectory and tends to converge to the reference angle~($90^\circ$). This reference angle corresponds to the elevation angle when the heights of both the transmitter and the receiver are the same. At higher UAV heights, the elevation angle starts at higher values and converges in a similar fashion to the reference elevation angle. 

\begin{figure}[!t]
	\centering
	\includegraphics[width=0.97\columnwidth]{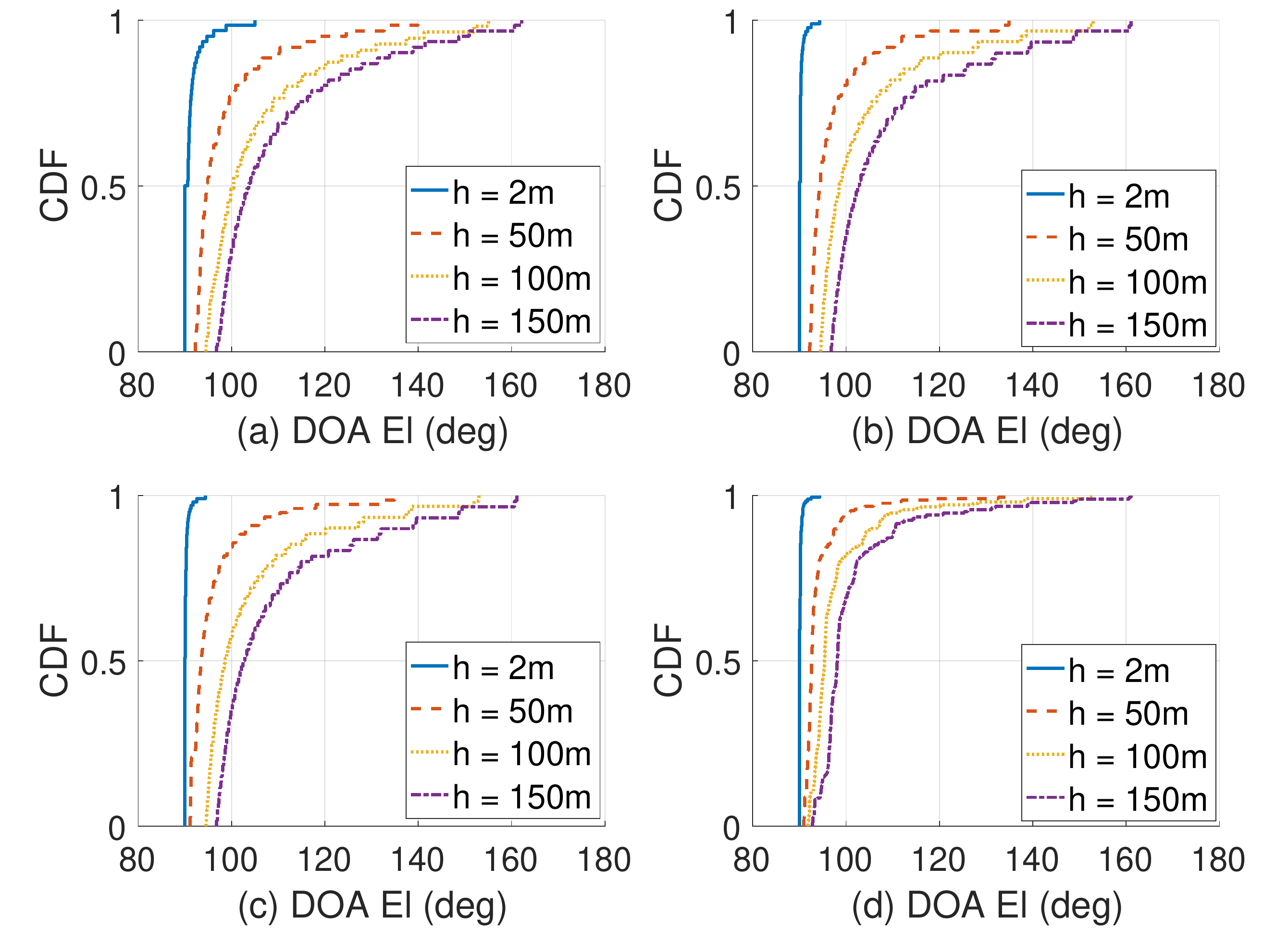}
	\caption{CDFs of DOA of MPCs $28$~GHz in the elevation plane for different propagation environments, (a) Over sea, (b) Rural, (c) Suburban, (d) Dense-urban. 
    }\label{Fig:CDF_AOA_El}   \vspace{-3mm}
\end{figure}

The CDF plots of the DOA in the elevation plane are shown in Fig.~\ref{Fig:CDF_AOA_El} for four different environments and four different UAV heights. The elevation DOA is observed to be dependent on the UAV height  in all the scenarios considered: higher the UAV height, higher the elevation angle.

\subsection{Direction of Departure of MPCs}
The DODs of the MPCs show similar behavior as the DOAs of the MPCs with smaller ranges as compared to arrival angles. The azimuth DODs of the MPCs at the transmitter side follow similar distribution as observed for DOAs with respective angular values, which are not repeated here due to space constraints. Rest of this subsection focuses on studying the elevation DODs of the MPCs.  




For DOD in the elevation plane we observe two distinct paths as shown in Fig.~\ref{Fig:AOA_AOD_El_overall_50m_suburban_urban}(c) for suburban scenario at UAV height of $150$~m. The upper one represents the GRC elevation angle and the lower one represents the LOS component. The GRC will be at higher elevation angle as compared to the LOS component at the start of the UAV trajectory. Both of them will converge to the reference angle along the UAV trajectory showing an increase in the elevation angle of the LOS component and a decrease in the elevation angle of the GRC along the UAV trajectory similar to what was observed in case of elevation DOA. There are additional non-persistent MPCs in case of dense-urban scenario with their respective elevation DODs shown in Fig.~\ref{Fig:AOA_AOD_El_overall_50m_suburban_urban}(d). The DODs of the non-persistent components are mostly around $90^\circ$ due to far-off location of scatterers from the UAV trajectory.    

\begin{figure}[!t]
	\centering
	\includegraphics[width=0.87\columnwidth]{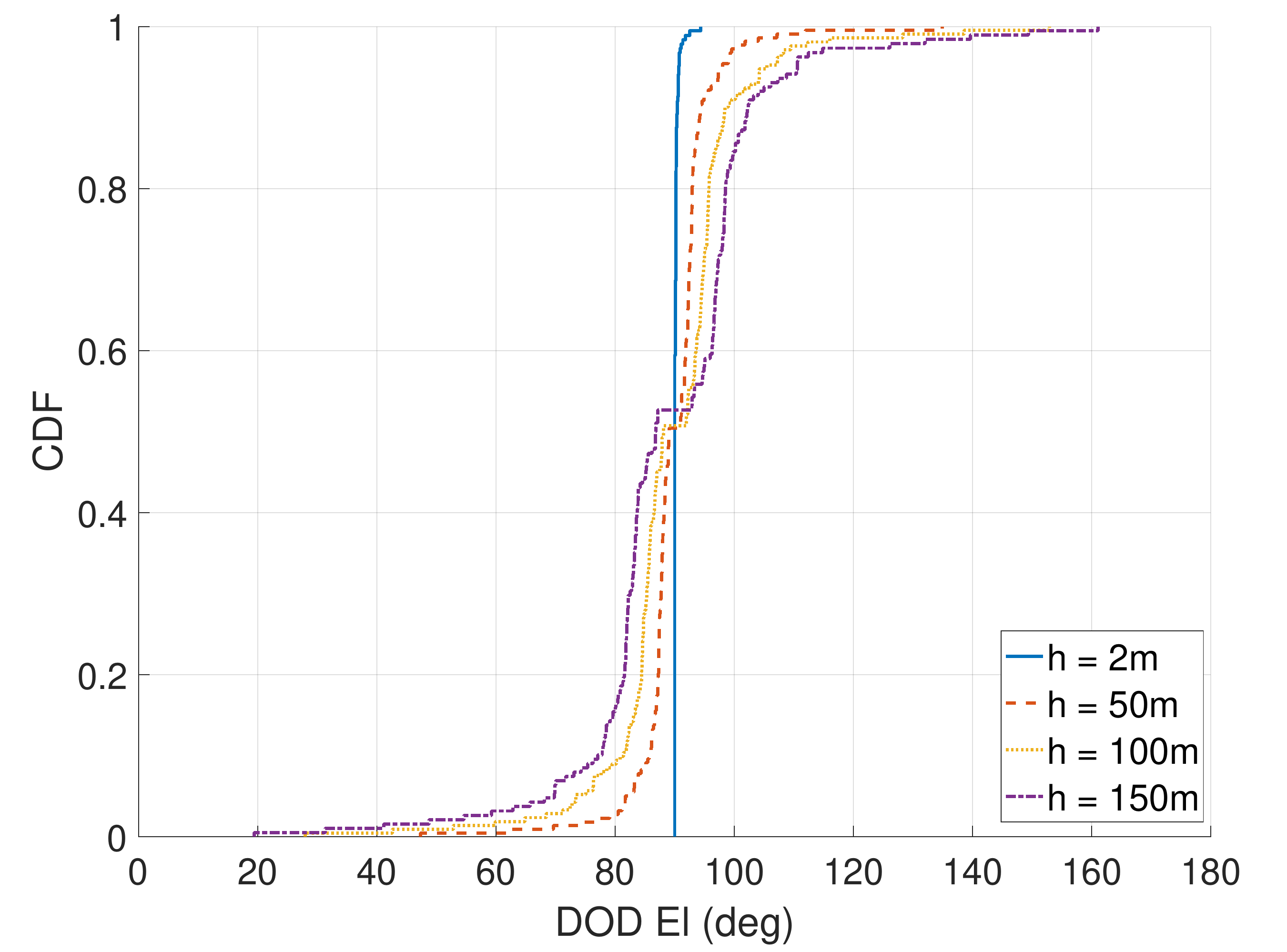}
	\caption{CDFs of DOD of MPCs at $28$~GHz in the elevation plane for dense-urban scenario.}\label{Fig:CDF_AOD_El_urban_alone}   \vspace{-4mm}
\end{figure}

The CDFs  of the DOD in the elevation plane are similar to those observed for DOA. For DOD in the elevation plane, a CDF plot for the dense-urban scenario is shown in Fig.~\ref{Fig:CDF_AOD_El_urban_alone}. It can be observed that at UAV height of $2$~m we have a constant DOD in the elevation plane, since both the UAV and the GS are aligned at same height. As the UAV height increases, we observe a corresponding increase in the spread of the elevation angle with a mean around $90^\circ$. This shape of the CDF is due to two different angles of the LOS and the GRC components as observed in Fig.~\ref{Fig:AOA_AOD_El_overall_50m_suburban_urban}(c)-(d).

\section{Conclusions}\label{Section:Conclusions}
In this work, we have explored the small scale and spatial characteristics of AG mmWave channels for different propagation environments of dense-urban, suburban, rural and over sea. It is observed that the MPCs reaching the receiver along the UAV trajectory can be divided into persistent and non-persistent MPCs. The characteristics of the persistent components will be dependent on the geometry of the setup and will not vary significantly along the UAV trajectory whereas the non-persistent components will be dependent mainly on the properties of the scatterers and follow a birth/death process along the trajectory of the UAV. Additionally, the mmWave AG channel characteristics are dependent on the relative height of the scatterers with respect to the UAV height. 

\bibliographystyle{IEEEtran}





\end{document}